\documentstyle[prl,aps,multicol,epsf,rotate]{revtex}
\def\beginwide{
        \end{multicols} \vspace*{-0.5cm} \noindent
        \rule{3.5in}{.1mm}\rule{.1mm}{5mm} \widetext \medskip }
\def\beginwidetop{
        \end{multicols} \vspace*{-0.5cm} \noindent
        \widetext \medskip }
\def\endwide{
        \hspace*{3.35in}~\rule[-5mm]{.1mm}{5mm}\rule{3.5in}{.1mm}
        \begin{multicols}{2} \vspace*{-1.0cm} \noindent }
\def\endwidebottom{
        \begin{multicols}{2} \vspace*{-1.0cm} \noindent }

\begin{document}
\tolerance = 10000
%
\title{Rayleigh loops in the random-field Ising model on the Bethe lattice}
\author{Francesca Colaiori, Andrea Gabrielli and Stefano Zapperi}
\address{INFM unit\`a di Roma 1, Dipartimento di Fisica,
Universit\`a "La Sapienza", P.le A. Moro 2
        00185 Roma, Italy}
\date{\today}
\maketitle
\begin{abstract}
We analyze the demagnetization properties of the random-field Ising
model on the Bethe lattice focusing on the beahvior near the disorder
induced phase transition. We derive an exact recursion relation for the
magnetization and integrate it numerically. Our analysis shows that 
demagnetization is possible only in the continous high disorder phase,
where at low field the loops are described by the Rayleigh law. In the
low disorder phase, the saturation loop displays a discontinuity 
which is reflected by a non vanishing magnetization $m_\infty$
after a series of nested loops. In this case, at low fields 
the loops are not symmetric and the Rayleigh law does not hold.
\end{abstract}
\pacs{PACS numbers:75.60.Ej, 75.60.Ch, 64.60.Ht, 68.35.Ct}

\begin{multicols}{2}
\section{Introduction}

A ferromagnetic material is characterized by a remanent magnetization
even at zero field. In several instances, however, it is convenient
to {\it demagnetize} the sample, bringing it to a state of zero 
magnetization at zero field.  In practice, 
this can done applying a slowly varying AC field, decreasing its amplitude 
after each cycle. In this way, the system explores a complex 
energy landscape, due to the interplay between structural disorder
and interactions, until it is trapped into a low energy minimum.
If the demagnetization process is performed adiabatically and thermal
effects do not play an important role, the 
demagnetized state is reproducible for a given realization
of the quenched disorder and can thus be used as a reference to
define magnetic properties.

The hysteresis loops at low fields, 
starting from the demagnetized state, are 
usually described by the  Rayleigh \cite{RAY-87} law:
when the field is cycled between $\pm H^*$ ,
the magnetization $m$ follows 
$m = (a+bH^*)H\pm b((H^*)^2-H^2)/2$, where
the signs $\pm$ distinguish the upper and lower branch of the loop.
Consequently the area of the loop scales with the peak field $H^*$ as
$W=4/3 b (H^*)^3$ and the response to 
a small field change, starting from the demagnetized  
state is given by $M^*=a(H^*)\pm b (H^*)^2$ \cite{Bertotti,Chikazumi}.
This law has been measured in a variety of materials, but
a few papers have reported significant deviations from the simple
quadratic law but no explanation has been provided \cite{BER-91}.

The current theoretical interpretation of this law is based on a 1942 paper
by N\'eel \cite{NEE-42}, who derived the law formulating the magnetization
process as the dynamics of a point (i.e. the position of a domain
wall) in a random potential. In this framework, the initial susceptibility 
$a$ is associated to reversible motions inside
one of the many minima of the random potential, while the hysteretic
coefficient $b$ is due to irreversible jumps between different valleys.
The main drowback of N\'eel theory relies in its purely phenomenological
nature, being based on a {\em zero dimensional} model
which does not include collective effects 
considered very important for the magnetization 
process \cite{ZAP-98,DUR-00,SET-01}.

In the past few years, the zero temperature 
random-field Ising model (RFIM)  has been used to describe 
the competition between quenched disorder and exchange interactions
and their effect on the hysteresis loop \cite{SET-93}.
In three and higher dimensions, the model shows a phase transition between a
continuous cycle for strong disorder and a discontinuous
loop, with a macroscopic jump, at low disorder. The two phases
are separated by a second order critical point, characterized
by universal scaling laws \cite{SET-93,DAH-96,PER-99} . 
A behavior of this kind is not restricted
to the RFIM but has also been observed in other models, with random
bonds \cite{VIV-94}, 
random anisotropies \cite{VIV-01} or vectorial spins \cite{DAS-99}. 
In addition, a similar disorder induced phase transition in the hysteresis loop
has been experimentally reported for a Co-Co0 bilayers \cite{BER-00}.

The RFIM is probably the simplest model including disorder
and exchange interactions that can be treated analytically.
The model has been solved exactly in one dimension \cite{SHU-96,SHU-00}
and on the Bethe lattice \cite{DHA-97,SAB-00,SHU-01}, 
while mean-field theory \cite{SET-93} 
and renormalization group \cite{DAH-96} have been used to
analyze the transition.
Recently the one dimensional solution of the model, has been 
generalized to obtain the complete demagnetization process and
to derive the Rayleigh loops \cite{DAN-01}. The RFIM does not
display a phase transition in one dimensions, while numerical
simulations indicate that the transition has an important effect
on the demagnatization process. In particular, in the low disorder
phase the discontinuity in the saturation curve prevents the 
magnetization to reach a demagnetized state \cite{DAN-01}
but this behavior has not been understood theoretically.  
It has been shown exactly that the RFIM displays a 
disorder induced phase transition on the Bethe lattice 
when the coordination number $z\geq 4$ \cite{DHA-97,SAB-00,SHU-01} and 
can thus be used to clarify the issue.

Here we generalize the analysis of 
Refs.~\cite{DHA-97,SAB-00,SHU-01} and \cite{DAN-01}
to obtain exact recursion relations for the demagnetization process
on the Bethe lattice and show that demagnetization is
only possible in the high disorder phase. In the low disorder
phase the remanent magnetization after a series of nested loops of
decreasing amplitude does not vanish but scales to zero as
the transition is approached. Furthermore, in the low disorder
phase the Rayleigh law is not obeyed and low field loops are
not symmetric. The Rayleigh law is instead recovered in the 
high disorder phase and the Rayleigh parameters $a$ and $b$ 
behave qualitatively as in $d=1$, displaying a peak in the
disorder.

The paper is organized as follows: in section II we describe
the model. In section III we recall the results obtained
in Refs.~\cite{DHA-97,SAB-00,SHU-01} and generalize them
for a series of nested loops. Section IV discusses the
effect of the phase transition on the magnetization and
section V analyzes the Rayleigh law. A brief discussion
of the perspectives is reported in section VI. Finally,
the complete derivation of the recursion relations is
reported in the Appendix.

\section{The model}

In this section we recall briefly a model used to describe  
hysteresis loops in magnetic materials \cite{SET-93}:
the ferromagnetic Random Field Ising Model (RFIM).  
This model is characterized by the Hamiltonian
\begin{equation}
{\cal H}=-J \sum_{\left<i,j\right>}s_i s_j-H\sum_{i}s_i-\sum_{i}h_i s_i
\label{hamiltonian}	
\end{equation}
where $J>0$, the $\sum_{\left<i,j\right>}$ 
is restricted on the pairs of nearest neighbors on 
a lattice of coordination number $z$, $s_i$ is the Ising spin on the site $i$,
$H$ is a homogeneous external field, and $h_i$ represents a 
quenched random field on the spin $s_i$ modelling 
the presence of lattice defects. The fields $\{h_i\}$   
are independently drawn from a symmetric distribution $\rho(h_i)$.
In the following the numerical results are referred to 
a Gaussian distribution with variance $R^2$.

In this paper we study the case of a Bethe lattice with a generic 
coordination number $z$. In particular we are interested in the case $z=4$ 
which is known to be the minimal case showing a disorder induced 
phase transition towards a discontinuous hysteresis loop.   

In order to mimic the microscopic spin dynamics,
we use the flipping rules used in 
Refs.~\cite{SET-93,DAH-96,PER-99,SHU-96,SHU-00,DHA-97,SAB-00,SHU-01}
obtained from the Glauber dynamics at temperature $T$
and with an external field of frequency $\omega$ taking 
the limit $T\rightarrow 0$ first and then $\omega\rightarrow 0$.
The basic rule of this $T=0$ dynamics is that the spins align with the 
local field:
\begin{equation}
s_{i}=\mbox{sign}(h_{i,eff})\,,
\label{sign}
\end{equation}
where the effective local field felt by the spin $i$ is 
\begin{equation}
h_{i,eff}=-\frac{\partial {\cal H}}{\partial s_i}=J\sum_{j\in n(i)}s_j+H+h_i
\label{eff_field}
\end{equation}
and the sum runs over the $z$ nearest neighbors of the site $i$. 

Note that though the model is defined through three external 
parameters $J,H,R$, the dynamics is determined by the two reduced 
quantities $H/R$ and $J/R$ only.
For the sake of simplicity, from now on we rename these two ratios
$H$ and $J$ respectively and consider $R=1$.
Given $H$ and $J$, we can write 
the probability $p_m(H)$ that a spin $i$, with $m$ ($0\le m\le z$) 
of its neighbors {\em up}, is also {\em up}.
This is given by the probability that $h_{i,eff}>0$:
\begin{equation}
p_{m}(H)=P(h_{i,eff}>0)=\int_{(z-2m)J-H}^{\infty}dh^{'}\rho(h^{'}) \,.
\label{pn}	
\end{equation}

\section{Hysteresis loops}

In general a change in the applied field 
$H$ produces a rearrangement of the spins, so that
each spin $i$ is stable being aligned with its effective field $h_{i,eff}$. 
It is important to note that each spin flip  modifies the effective 
field on the nearest neighbors and sometimes
generates an avalanche of spin flips through the lattice.
In the following we will consider the case of a slowly varying external field: 
its value is kept constant until the next metastable state is reached. 
Two important properties of the $T=0$ dynamics are (i) the 
Abelian property -- the stable state after an avalanche does not depend on 
the order in which the spins flip --, and (ii) the return point memory -- when 
the field is changed adiabatically the stable state only depends on 
the point where the field was last reversed. These two properties can
be used to obtain exactly the shape of the hysteresis loops.
We first recall the derivation of the saturation curve and those
of the first minor loops and then procede with the general analysis
of minor loops.

\subsection{Saturation loop}

When the external field $H$ is cycled from $-\infty$ to $+\infty$ and back
the magnetization describe the saturation loop.
The key quantity describing the lower half of the saturation loop is the
conditional probability $U_0(H)$ \cite{DHA-97} defined as 
the probability that a given spin flips before 
a fixed nearest neighbor, conditioned to this neighbor being down.
The probability $U_0(H)$ satisfies the following equation \cite{DHA-97}:
\begin{equation}
U_{0}(H)=\sum_{m=0}^{z-1} {z-1 \choose m} [U_{0}(H)]^{m}
[1-U_{0}(H)]^{z-1-m} p_{m}(H)	 \,.
\label{U02}
\end{equation}
It can be shown that
the probability that a spin is {\em up} at external field $H$ is \cite{DHA-97}
\begin{equation}
p(H)=\sum_{m=0}^{z}{z \choose m} [U_{0}(H)]^{m}
[1-U_{0}(H)]^{z-m} p_{m}(H) \,.	
\end{equation}
The related magnetization is given by 
$m_l(H)=1-2\,p(H)$, which describes the lower half of the hysteresis loop. 
The upper half of the hysteresis loop can then be obtained by symmetry
(i.e. $m_u(H)=m_l(-H)$).

\subsection{First minor loops}

If the external field $H$ is raised from $-\infty$  to a finite value $H_0$ 
(i.e. we are on the lower half of the major loop) and then it is reversed, 
the magnetization describes the upper half of a minor hysteresis loop.
When the field is reversed from $H_0$ to $H_1<H_0$ we define
the conditional probability $D_1(H_1)$ for a spin to be {\em down} before 
a fixed nearest neighbor, conditioned to this neighbor being {\em up}. 
The probability $D_1(H_1)$ satisfies the following equation \cite{SHU-01}
\beginwide
\begin{equation}
D_{1}(H_{1})=\sum_{m=0}^{z-1} {z-1 \choose m} [U_{0}(H_0)]^{m} \left\{
[1-U_{0}(H_0)]^{z-1-m}[1- p_{m+1}(H_0)]+
[D_{1}(H_1)]^{z-1-m}[p_{m+1}(H_0)- p_{m+1}(H_1)]	 \right\} \,.
\label{d1}
\end{equation}
The related probability that a spin is {\em up} at $H_1$ is 
\begin{equation}
p(H_1)=p(H_0)-\sum_{m=0}^{z}{z \choose m} [U_{0}(H_0)]^{m}
[D_{1}(H_1)]^{z-m} [p_{m}(H_0)-p_m(H_1)]	\,.
\end{equation}\endwide
Equation (\ref{d1}) holds as long as $H_1$ is larger than $H_0-2J$ 
\cite{SHU-01}. In one dimension it has been shown \cite{SHU-00} 
that at $H =H_0-2J$
the upper half of minor loop merges with the upper half of the major loop
with the same local slope.
This proof can be extended to the case of a Bethe lattice as long
as the saturation loop is continuous \cite{SHU-01}. 
The case in which the saturation loop displays a discontinuity is discussed 
below for the case $z=4$.

\subsection{General formula for nested loops}

The method used to find $U_0$ and $D_1$ can be generalized to obtain a complete
characterization of all minor loops.
In particular we are interested in {\em nested}
minor loops, since they are directly related to the demagnetization process
of the disordered ferromagnet.
Nested loops is defined as follows: 
after having reached $H_1$, we reverse again the field
increasing its value up to $H_2\le H_0$ (lower half of the first minor loop).
This process is then iterated in a sequence of fields
$H_{2n}\in [H_{2n-1},H_{2n+1}]$
and $H_{2n+1}\in [H_{2n},H_{2n+2}]$
with $n\ge 1$, where $H_{2n}$ and $H_{2n+1}$ refer 
to the final value of the field $H$
in the lower half of the $n^{th}$ minor loop, and for the 
upper half  of the $(n+1)^{th}$ minor loop respectively.
The generalizations of $U_0$ to the minor loops is called $U_{2n}(H_{2n})$
while the generalization of $D_1(H_1)$ is called $D_{2n+1}(H_{2n+1})$.
In what follows, we simply indicate $U_{2n}(H_{2n})$ with $U_{2n}$
and $D_{2n+1}(H_{2n+1})$ with $D_{2n+1}$.

Since $H_{2n}<H_{2n-2}$, the set of spins contributing to 
$U_{2n}$ will be a subset of those contributing to $U_{2n-2}$,
so that we can write 
\begin{equation}
U_{2n}=U_{2n-2}-\eta_{2n-1}+\eta_{2n}	
\label{U2n}
\end{equation}
where $\eta_{2n-1}$ represents the fraction of spins that 
were {\em up} at $H_{2n-2}$ 
before their fixed nearest neighbor and {\em down} at $H_{2n-1}$,
while $\eta_{2n}$ is the fraction of the set contributing to $\eta_{2n-1}$
which flip {\em up} again at $H_{2n}$.
The explicit derivation of $\eta_{2n-1}$ and $\eta_{2n}$ is a little
involved and it is thus discussed in the appendix.

The magnetization at $H=H_{2n}$ is as usual obtained 
as $m_{2n}\equiv m(H_{2n})=1-2\,p(H_{2n})$, where 
the probability $p(H_{2n})$ that a spin is {\em up} at $H_{2n}$ 
is given by the probability $p(H_{2n-1})$ that it was already {\em up} 
at $H_{2n-1}$ summed to the probability to 
flip {\em up} when the field goes from $H_{2n-1}$ to $H_{2n}$:
\beginwide
\begin{equation}
p(H_{2n})=p(H_{2n-1})+\sum_{m=0}^{z}{z \choose m} [U_{2n}]^{m}
[D_{2n-1}]^{z-m} [p_{m}(H_{2n})-p_m(H_{2n-1})]	\,.
\end{equation}
\endwide
The generalization of $D_1$ to nested minor loops is called 
$D_{2n+1}$ and analogously to $U_{2n}(H_{2n})$ is given by
\begin{equation}
D_{2n+1}=D_{2n-1}-\zeta_{2n}+\zeta_{2n+1}\,,
\label{D2n+1}
\end{equation}
where $\zeta_{2n}$ is the fraction of spins that 
were {\em down} at $H_{2n-1}$ before their fixed 
nearest neighbor and {\em up} at $H_{2n}$ 
and $\zeta_{2n+1}$ is the fraction of the set of spins
contributing to $\zeta_{2n}$ which
flip {\em down} again at $H_{2n}$.
The exact expression for the fractions $\zeta_{2n}$ and $\zeta_{2n+1}$ are 
reported in the appendix.

The magnetization at $H=H_{2n+1}$ is 
$m_{2n}\equiv m(H_{2n})=1-2\, p(H_{2n+1})$, where
$p(H_{2n+1})$ is the probability for a spin to be {\em up} at $H_{2n+1}$.
This probability can be written as the analogous probability
at $H=H_{2n}$ minus the probability to flip {\em down} between $H_{2n}$
and $H_{2n+1}$:
\beginwide
\begin{equation}
p(H_{2n+1})=p(H_{2n})-\sum_{m=0}^{z}{z \choose m} [U_{2n}]^{m}
[D_{2n+1}]^{z-m} [p_{m}(H_{2n})-p_m(H_{2n+1})]\,.	
\end{equation}
\endwide

In principle, an arbirary series of nested loop can be obtained   
solving the recursion relation for $U_{2n}$, $D_{2n+1}$, and
using the result to obtain the magnetization. A similar procedure
was used in $d=1$ to obtain a closed expression for the magnetization 
along the demagnetization curve. This is not possible for the Bethe
lattice where an explicit solution for the problem is not available and  
one should resort to a numerical integration.

\section{Disorder induced phase transition and demagnetization}

Previous numerical studies of the zero temperature dynamics of the 
RFIM on a regular lattice in finite dimension \cite{SET-93,PER-99} have 
shown that in $d\geq 3$ the system 
shows a phase transition from a strong disorder phase to a weak 
disorder phase, separated by a second order critical point 
(in $d=2$ the presence of a phase transition is still controversial). 
In the strong disorder phase the major hysteresis loop is continuos, 
whilst in the weak disorder phase the loop shows a macroscopic jump in 
the magnetization at a critical value of the field. 
Note that if we use the reduced parameters $J/R$ and $H/R$, 
fixing $R=1$, strong disorder corresponds to small values of $J$
and the phase transition will be charcterized by a critical value $J_c$
of the exchange coupling $J$. 
On the Bethe lattice, a phase transition 
is observed for coordination numbers $z\ge 4$, while for $z\le 3$ one has
only the strong disorder phase \cite{DHA-97}.
The presence of the phase transition has strong implications on the
possibility of demagnetizing the system. 
In particular in the weak disorder phase phase it is not possible to 
demagnetize the system through an oscillating external field
with decreasing amplitude.

Before we specialize to the case $z=4$, let us note that for $H=J$ 
one has $p_{z-1-m}(J)=1-p_m(J)$. This allows, after a little algebra, 
to show that for  $H=J$, $U_0(J)=1/2$ is a solution for any $J$, 
any $z$ and any random field distribution. 
For $z=4$ Eq.~(\ref{U02}) is a cubic equation in $U_0$, 
with coefficients depending on $H$ and $J$ through the $p_i$.
In order to find the critical point it is enough to find the value of
$H$ and $J$ corresponding to a triple solution of the equation.
Implementing this requirement for any simmetric density function
of the disorder, one finds $H_c=J_c$, where 
$J_c$ satisfies the equation $p_0(J)+p_1(J)=1/3$.
For a Gaussian distribution of the disorder, this translates in  
the following implicit equation for $J_c$:
\begin{equation}
\mbox{Erf}(J_c)+\mbox{Erf}(3J_c)=1/3
\label{Jc}
\end{equation}
resulting in  $J_c=0.56140099587319...$, in accordance with
the result quoted Ref.~\cite{SHU-01}.

Above the transition ($J>J_c$ or weak disorder phase) the hysteresis loop
becomes discontinuous \cite{DHA-97}.
In fact at $J=J_c$ the and $H=J_c$ the susceptibility $\partial m/\partial H$
diverges, and for $J>J_c$ one observes a discontinuity whith a spinodal
singularity.
At this point one can measure a gap $\Delta m$ in the magnetization. 
It is easy to show, through an expansion 
of Eq.~(\ref{U02}) to the lowest order in $J-J_c$ around $J_c$, 
that for $J\rightarrow J_c^+$ it is 
\begin{equation}
\Delta m\sim (J-J_c)^\beta
\label{gap}
\end{equation} 
with $\beta=1/2$ as in the mean field case.
The analytical derivation of this result can be easily sketched as follows:
first of all the three solutions of Eq.~(\ref{U02}) at $H=J$ can be found
explicitly. 
They are$U_0^{(a)}=1/2$ and 
$U_0^{(b),(c)}=1/2(1\pm \sqrt{(1-3(p_0+p_1))/(1-3p_1+p_0)})$,
where, for $H=J$, $(1-3(p_0+p_1))>0$ only for $J>J_c$ and $=0$ at $J=J_c$.
The gap can be measured by $\Delta U\equiv 
\mid U_0^{(b)}-U_0^{(c)}\mid$, and it is simple
to show that $\Delta U\sim (J-J_c)^{1/2}$, which gives Eq.~(\ref{gap})
considering that to the lowest order $\Delta m\sim\Delta U$.
Actually, the position of the gap in the mgnetization of the 
major loop of the hysteresis cicle is located at $H_g>J$, which could give
corrections to the previous result.
However, one can show that $(H_g-J)\sim (J-J_c)$, implying corrections
to the previous value of the gap of the same order in $(J-J_c)$. Then this 
correction does not alter the found scaling behavior.

We recall that usually it is possible 
to demagnetize a material by applying a slowly oscillating external 
field appropriately chosen.
In practice, this corresponds to a series of nested loops, 
starting from the completely magnetized situation (e.g. $m=-1$ and 
$h\rightarrow -\infty$) and then applying in succession the fields 
$H_0=J$, $H_1=-H_0(1-\varepsilon)$, ..., 
$H_{2n+1}=-H_{2n}(1-\varepsilon)$ in the
limits $\varepsilon \rightarrow 0^+$ and $n\rightarrow +\infty$ \cite{DAN-01}.
This sequence, for $J<J_c$, leads to a completely demagnetized state. 
This is due to two fundamental properties of the strong disordered phase:
(i) the ``return point memory'' property which has been defined in
the previous paragraph; (ii) the fact that, if we are on a certain point 
of the saturation loop (e.g. at $H=J_0$ on the lower half) and invert the field
to $H_1$ the system meets the other half on the saturation loop 
if $H_1=H_0-2J$.
This implies that in order that the first minor loop is symmetric with respect 
to the origin $H=0$ and $m=0$ without touching the saturation curve we 
have to start from $H_0=J$ on the lower half of the saturation
loop (or equivalently from $h_0=-J$ on the upper half). 

\begin{figure}
\narrowtext\centerline{\epsfxsize\columnwidth\epsfbox{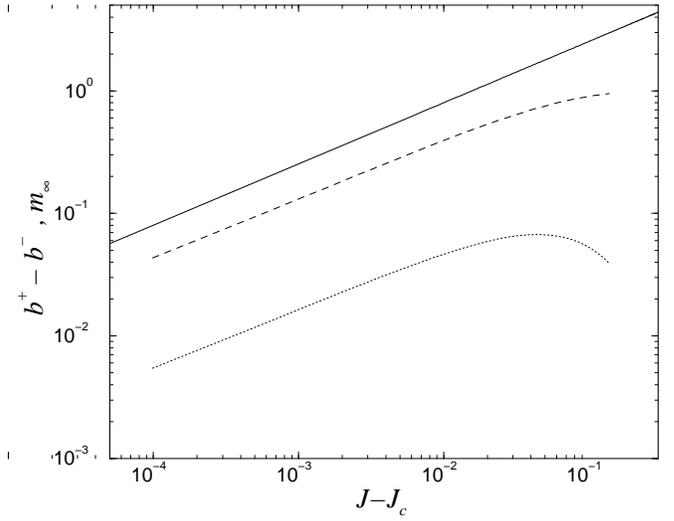}}
\caption{Final magnetization $m_\infty$ and $\Delta b$ as a function 
of $J-J_c$. The solid line has a slope $1/2$.}
\label{log}
\end{figure}

At $J>J_c$ the demagnetization process is no more possible, because the 
discontinuity prevents minor loops to be symmetric with respect 
to the origin of the axes.
In fact there is now an inaccessible region of the plane $(H,m)$ around
the origin \cite{DAN-01}. However, the field succession described above
still provides a well defined procedure (apart the broken simmetry
$m \rightarrow -m$ and $H\rightarrow -H$) to 
minimal possible residual magnetization,  that we denote $m_{\infty}$,
at $H=0$.  Following this procedure, by numerically integrating 
Eqs.~(\ref{U2n},\ref{D2n+1}) with $\varepsilon=10^{-3}$,
we find that $m_{\infty}$ displays the same 
scaling behavior of the gap in the saturation loop
as $J$ approaches $J_c$ from above: 
$m_{\infty} \propto (J-J_c)^{1/2}$ (see fig(\ref{log})). 
Notice that in numerical simulations in the $d=3$ RFIM 
\cite{DAN-01} $m_{\infty}$ was found to coincide, in the
low disorde phase, with the saturation magnetization $m_0(0)$.

\section{Rayleigh law}

The Rayleigh law describes the hysteresis behavior at
low field in a vast class of materials. Exact values for
the Rayleigh parameters have been obtained for the 
one dimensional RFIM \cite{DAN-01}, where the 
initial susceptibilty $a$ and the hysteretic coefficient
$b$ both display a peak in the disorder $R$. A similar
behavior is observed in simulation for $d=2,3$ but only
in the high disorder phase, while in the low disorder phase
$a$ and $b$ are not defined.

\begin{figure}
\narrowtext\centerline{\epsfxsize\columnwidth\epsfbox{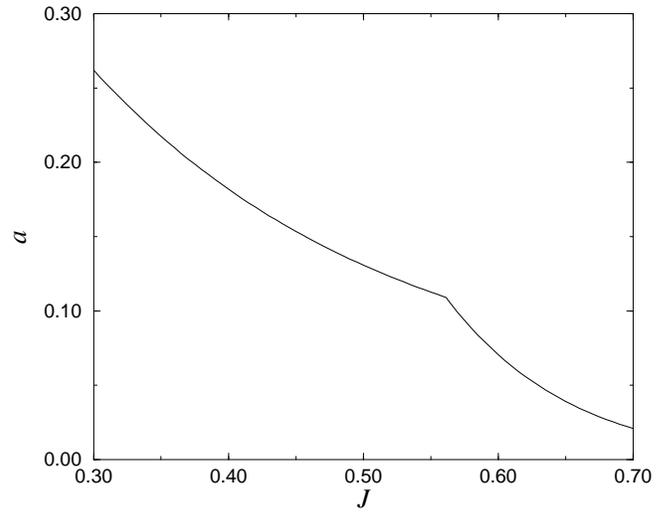}}
\caption{Susceptibility $a$ as a function of the 
exchange coefficient $J$.}
\label{a}
\end{figure}

\begin{figure}
\narrowtext\centerline{\epsfxsize\columnwidth\epsfbox{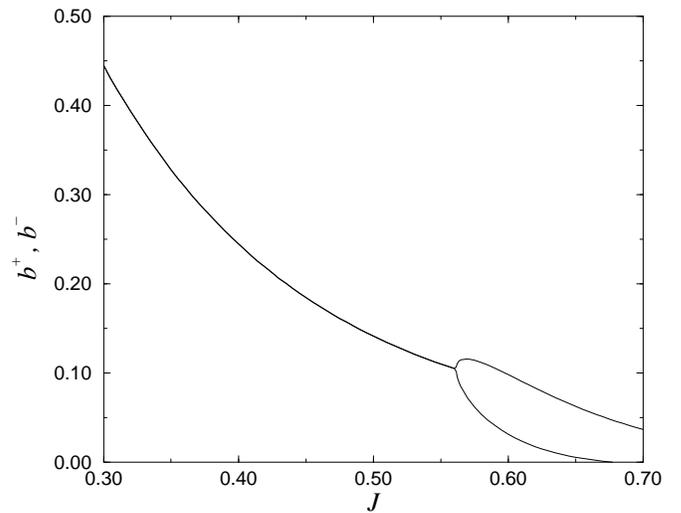}}
\caption{Hysteretic coefficient $b$ as a function of the 
exchange coefficient $J$.}
\label{bb}
\end{figure}

In the case of the Bethe lattice, we could not obtain an 
explicit expression of the Rayleigh parameters even in the
high disorder phase. We thus resort to numerical integration
and analyze the demagnetization curves close to $H=0$.
We estimate the susceptibility $a$ and the 
hysteretic coefficient $b$ using a linear fit of $m_{2n}/H_{2n}$ vs $H_{2n}$.
According to the Rayleigh law for $n\to\infty$ we have
$m_{2n}/H_{2n}=a+bH_{2n}$, and similarly 
for negative fields $m_{2n+1}/H_{2n+1}=a-bH_{2n+1}$.
The values of $a$ and $b$ as a function of the exchange 
coupling $J$ are shown in Figs~\ref{a} and \ref{bb}. When
plotted as a function of the disorder $R$, $a$ and $b$ show
a peak in the high disorder phase, in agreement which the results
on Euclidean lattices.
 
\begin{figure}
\narrowtext\centerline{\epsfxsize\columnwidth\epsfbox{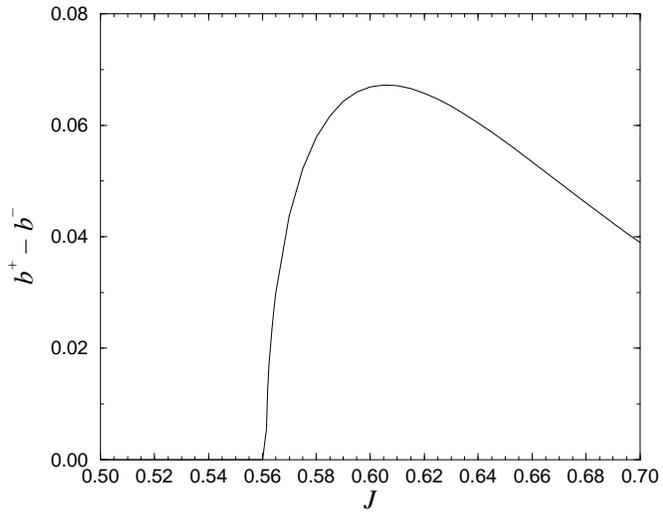}}
\caption{The deviations from the Rayleigh law are measured
by $\Delta b =b^+-b^-$.}
\label{b-b}
\end{figure}

In the low disorder phase the demagnetization curve is not
symmetric with respect to $H=0$ and consequently the Rayleigh
law does not hold. In particular, we can define two values
for the coefficient $b$: 
\[
m_{2n}/H_{2n}=a+b^+H_{2n}~~\mbox{and}\]
\begin{equation}
m_{2n+1}/H_{2n+1}=a-b^- H_{2n+1}.
\end{equation}
The values of $a$, $b^+$, and $b^-$ as a function of the exchange 
coefficient $J$ are shown in fig (\ref{a},\ref{bb}); also the difference 
$\Delta b =b^{+}-b^{-}$ is shown in fig (\ref{b-b}), 
Again, as  $J\rightarrow J_c^{+}$, $\Delta b$ approaches $0$ as 
$(J-J_c)^{1/2}$ (see fig(\ref{log})). 

\section{discussion}

In conclusion, the present analysis allows to clarify the
role of a disorder induced phase transition on the demagnetization
properties of a ferromagnet. In particular, we find that in the
low disorder phase the jump in the saturation curve gives rise
to an inaccessible region in the ($m$,$H$) plane close to $H=0$
and $m=0$. Even after an infinitesimally fine series of nested
loop the final magnetization $m_\infty$ does not vanish. Approaching the
transition, however,  $m_\infty$ scales to zero with an exponent $1/2$,
which is the same as the one controlling the size jump in the
saturation curve. While this value could be an artifact of the Bethe
approximation, the impossibility to demagnetize the system in the low
disordered phase has already been observed in three dimensional
numerical simulations \cite{DAN-01,CAR-01}.

\section*{acknowledgments}
This work is supported by the INFM PAIS-G project on ``Histeresis
in disordered ferromagnets''. We thank L. Dante, G. Durin and A. Magni
for useful discussions and G. Caldarelli for his warm encouragement.

\section*{appendix}
Here we derive the expressions for $\eta_{2n-1}$,
$\eta_{2n}$, $\zeta_{2n}$ and $\zeta_{2n+1}$ appearing in Eqs.~(\ref{U2n})
and (\ref{D2n+1}). We first note that these quantities can be 
defined in a recursive way. In particular, $\eta_{2n-1}$ 
represents the fraction of the set of spins which, at $H_{2n-2}$, contribute 
to $U_{2n-2}$, but are {\em down} at $H_{2n-1}$. To obtain the weight
associated to this fraction, consider a spin $i$ with a 
given neighbor $j$ kept  {\em down} (i.e. the spin $j$ is 
conditioning the probabilities). 
When the spin $i$ flips {\em down} at $H_{2n-1}$, apart from $j$
all the other neighbors can be either {\em up} or {\em down}. 
Consider for instance the case in which $m$ of these 
neigbors are {\em up} and $z-1-m$ {\em down}.
Under these conditions, the spin $i$ flips {\em down} if its effective field 
is positive at $H_{2n-2}$ and negative at $H_{2n-1} < H_{2n-2}$.
The associated contribution to $\eta_{2n-1}$ is then given by
\begin{equation}
[U_{2n-2}]^{m}
[D_{2n-1}]^{z-1-m} [p_{m}(H_{2n-2})-p_{m}(H_{2n-1})]\,
\label{eta-m}
\end{equation}
where $[U_{2n-2}]^{m}[D_{2n-1}]^{z-1-m}$ is the 
probability that, at the moment at which the spin $i$ flips {\em down}, 
$m$ given neighbors are {\em up} and $z-1-m$ (other than $j$) are {\em down},
and $[p_{m}(H_{2n-2})-p_{m}(H_{2n-1})]$ is the probability that the spin
$i$, having $m$ {\em up} neighbors, is {\em up} at $H_{2n-2}$ 
but not at $H_{2n-1}$. To obtain $\eta_{2n-1}$, we have first to multiply
Eq.~(\ref{eta-m})  by a combinatorial factor 
$z-1 \choose n$, taking into account all the equivalent choices
of the site $j$, and then sum over $m$ from $0$ to $z-1$. The
results reads:
\beginwide
\begin{equation}
\eta_{2n-1}=\sum_{m=0}^{z-1} {z-1 \choose m} [U_{2n-2}]^{m}
[D_{2n-1}]^{z-1-m} [p_{m}(H_{2n-2})-p_{m}(H_{2n-1})].	
\label{eta}
\end{equation}
\endwide

An analogous procedure can be implemented to derive $\eta_{2n}$, the
fraction of the set of spins contributing to
$\eta_{2n-1}$ which flip back {\em up} at $H_{2n}$. 
Using a derivation similar to the one discussed above, we obtain
\beginwide
\begin{equation}
\eta_{2n}=\sum_{m=0}^{z-1} {z-1 \choose m} [U_{2n}]^{m}
[D_{2n-1}]^{z-1-m} [p_{m}(H_{2n})-p_{m}(H_{2n-1})].
\end{equation}
\endwide
The quantities $\zeta_{2n}$ and $\zeta_{2n+1}$ can be obtained 
proceeding as in the evaluation of $\eta_{2n-1}$ and $\eta_{2n}$,
noticing that  in this case
the fixed neighbor $j$ (conditioning the probabilities) has
to be kept kept {\em up}.
Moreover the spin $i$
must flip from {\em down} to {\em up} at $H_{2n}$, in order to contribute
to $\zeta_{2n}$, and then flip back
{\em down} at $H_{2n+1}$ in order to contribute also to $\zeta_{2n+1}$. 
The final results reads
\beginwide
\begin{equation}
\left\{
\begin{array}{lll}
\zeta_{2n}&=&\sum_{m=0}^{z-1} {z-1 \choose m} [U_{2n}]^{m}
[D_{2n-1}]^{z-1-m} [p_{m+1}(H_{2n})-p_{m+1}(H_{2n-1})]\\
\\
\zeta_{2n+1}&=&\sum_{m=0}^{z-1} {z-1 \choose m} [U_{2n}]^{m}
[D_{2n+1}]^{z-1-m} [p_{m+1}(H_{2n})-p_{m+1}(H_{2n+1})]\,.	
\end{array}
\right.
\label{zeta}
\end{equation}
\endwide

\begin{figure}
\narrowtext\centerline{\epsfxsize\columnwidth\epsfbox{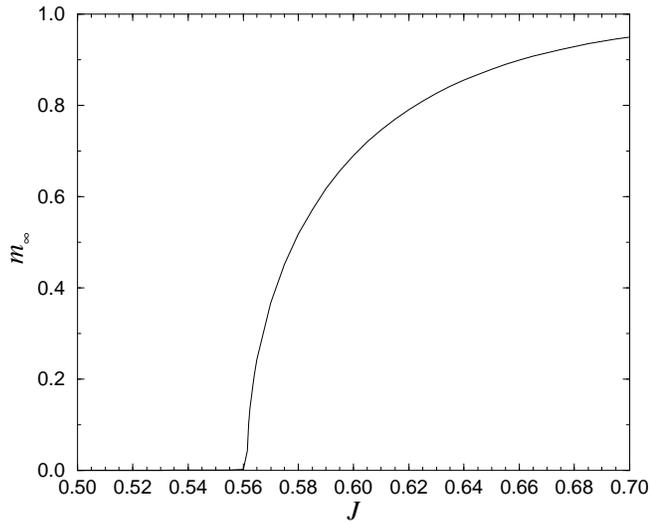}}
\caption{Final magnetization $m_\infty$ as a function of the 
exchange coefficient $J$.} 
\label{c}
\end{figure}
\begin{figure}
\narrowtext\centerline{\epsfxsize\columnwidth\epsfbox{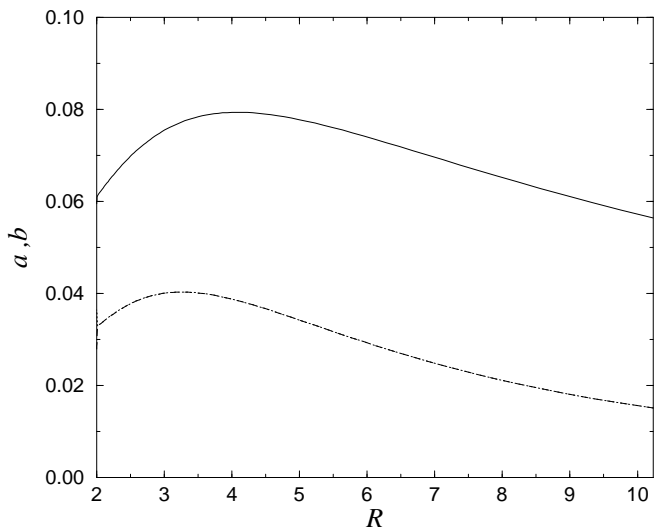}}
\caption{The Rayleigh parameters (solid line is $a$, dashed 
line is $b$) as a function of the disorder
width $R$ for $R>R_c=1/J_c$.} 
\label{r}
\end{figure}




\end{multicols}
\end{document}